# Reconfiguration of Brain Network between Resting-state and Oddball Paradigm

**Fali Li[1], Chanlin Yi[1], Yuanyuan Liao[1], Yuanling Jiang[1], Yajing Si[1], Limeng Song[1], Tao Zhang[1], Dezhong Yao[1,2], Yangsong Zhang[1,2,3], Zehong Cao[4] and Peng Xu[1,2]**

[1]The Clinical Hospital of Chengdu Brain Science Institute, MOE Key Lab for NeuroInformation, University of Electronic Science and Technology of China, Chengdu, 611731, China
[2]School of Life Science and Technology, Center for Information in Medicine, University of Electronic Science and Technology of China, Chengdu, 611731, China
[3]School of Computer Science and Technology, Southwest University of Science and Technology, Mianyang, 621010, China
[4]Centre for Artificial Intelligence, Faculty of Engineering and IT, University of Technology Sydney, NSW, Australia

Corresponding author: Yangsong Zhang (zhangysacademy@gmail.com), Peng Xu (xupeng@uestc.edu.cn).

This work was supported by the National Key Research and Development Plan of China (#2017YFB1002501), the National Natural Science Foundation of China (#61522105, #61603344, #81401484, and #81330032), the Open Foundation of Henan Key Laboratory of Brain Science and Brain-Computer Interface Technology (No. HNBBL17001), and the Longshan academic talent research supporting program of SWUST (#17LZX692).

**ABSTRACT** The oddball paradigm is widely applied to the investigation of multiple cognitive functions. Prior studies have explored the cortical oscillation and power spectral differing from the resting-state conduction to oddball paradigm, but whether brain networks existing the significant difference is still unclear. Our study addressed how the brain reconfigures its architecture from a resting-state condition (i.e., baseline) to P300 stimulus task in the visual oddball paradigm. In this study, electroencephalogram (EEG) datasets were collected from 24 postgraduate students, who were required to only mentally count the number of target stimulus; afterwards the functional EEG networks constructed in different frequency bands were compared between baseline and oddball task conditions to evaluate the reconfiguration of functional network in the brain. Compared to the baseline, our results showed the significantly ($p < 0.05$) enhanced delta/theta EEG connectivity and decreased alpha default mode network in the progress of brain reconfiguration to the P300 task. Furthermore, the reconfigured coupling strengths were demonstrated to relate to P300 amplitudes, which were then regarded as input features to train a classifier to differentiate the high and low P300 amplitudes groups with an accuracy of 77.78%. The findings of our study help us to understand the changes of functional brain connectivity from resting-state to oddball stimulus task, and the reconfigured network pattern has the potential for the selection of good subjects for P300-based brain-computer interface.

## I. INTRODUCTION

In recent years, P300 component, a positive deflection in the human event-related potential (ERP), has been widely investigated by using various neuroimaging technologies, such as electroencephalogram (EEG) and functional magnetic resonance imaging (fMRI). The P300 is usually evoked by the presentation of target stimulus in the oddball paradigm, in which two types of stimulus (i.e., target and standard) are presented to subjects in a random order. During the stimulus experiments, subjects are required to provide the responses to multiple target stimuli (i.e., counting the number or pressing a button once they notice the appearance of target stimulus) as correctly and quickly as possible, while omit standard ones. The P300 has been demonstrated to correlate with various cognitive functions (e.g., attention and decision making) [1-4]. For example, the P300 is regarded as a biomarker to evaluate to which degree the information can be processed by the brain in individual subjects. Actually, the P300 has been also widely used in multiple aspects, such as clinical diagnosis [5, 6], cognitive neuroscience [1, 7], and brain-computer interface (BCI) [8-11]. In terms of the practical applications, the P300 is quantitatively depicted with amplitude and latency. The amplitude is measured depended on the grand-averaged ERP, which is definitely defined as the largest positive peak within the time window of 300-500 ms [12]; while, the latency is the interval between stimulus onset and time point of largest peak, which are both demonstrated to show huge variability across individual subjects [7, 13]. However, it is still unclear that the mechanism underlying the P300 variability across individual subjects.

Multiple brain regions, such as prefrontal, frontal, and parietal, have been demonstrated to play important roles in the generations of P300 components [12, 14-16], while lesions in the certain part of these areas, such as frontal lobe, are clarified to result in the deficits in P300 components, such as the decreased amplitude [14, 17]. In essence, the specified brain activities in different frequency bands (i.e., delta, theta, and alpha bands) were demonstrated to closely relate to the cognitive processes and the P300 components [18-21]. Our brain cortex and functions are working within a large-scale complex network, and the information can be thereby processed between brain areas that are specialized, spatially distributed but functionally coupled [22]. The network neuroscience raises a 'bridge' between the brain networks and cognitive architectures [23]. For example, the higher intelligence scores is demonstrated to correspond to a more efficient information transfer in the brain [24]; and the larger brain responses investigated in steady-state visually evoked potential based studies are also clarified to be correlated to the better network topology structures [25, 26]. In fact, the generation of P300 is also attributed to the inter-regional activity which is contributed from multiple brain areas [27-30]. Our previous studies on either resting-state [29] or oddball stimulus task [28] have consistently demonstrated the close relationships between the couplings of prefrontal and parietal/occipital lobes and P300 components, which could also be validated by the synchronized brain activity that originated from the frontal-posterior network [30]. Herein, the brain network analysis is widely adopted to investigate the neural mechanism that accounts for the generations of P300; and multiple methods, including the coherence and Granger causality, are usually applied in brain connectivity studies [31-34].

Our brain is not idle, even in a resting-state condition [35]. The spontaneous brain activity at rest may reflect the potential capacity of the brain in efficiently processing the information during tasks [31, 36, 37]. Previous studies have demonstrated the possibility of resting-state EEG signals in studying the pathophysiological mechanisms of different diseases, such as epilepsy [38-44]. In fact, not only the spontaneous EEG parameters (i.e., spectral power and network efficiency) [29, 45], but the parameters derived from task EEG signals are consistently demonstrated to be helpful to deepen our knowledge of the generations of P300 [28, 30]. However, previous studies merely focus on single modality of either resting-state or stimuli task to investigate the P300, few studies have considered the relationship between these two brain states, and most importantly, it is still left unveiled how the brain reorganizes from a resting state to fulfill the needs of P300. Therefore, in this study, we assumed that the efficient reconfigurations from a resting-state to an oddball task in the brain are highly associated with the P300-related information processing. We then collected the resting-state and P300 task EEG datasets of 24 subjects who participated in our oddball P300 experiments. By statistically comparing the EEG networks in different frequency bands between the resting-state and P300 task conditions, we evaluated the updates of functional connectivity in the progress of brain reconfiguration from resting-state to P300 task, and finally investigated the underlying relationships between the reconfigured networks and P300 components.

## II. Materials and methods

### A. PARTICIPANTS
Twenty-four healthy right-handed postgraduate students (males, age range of 22-27 years) were compensated financially to participate in our experiment after providing their signed written informed consent. Of note, none of them had a history of substance abuse, and a personal or family history of psychiatric or neurological disease. This experiment has been approved by the Institution Research Ethics Board of University of Electronic Science and Technology of China.

### B. EXPERIMENTAL PROCEDURES
In the preparation stage, all subjects were asked to remain relaxed, and focus their attention on the center of the computer screen. They were also required to refrain from extensive head motion, to avoid moving their body during the whole experiments. Then, we collected EEG signals in a standard protocol which consists of a 4-minute eyes-closed resting-state condition and following 3-times P300 tasks. The procedures of the experimental design are illustrated in Figure 1, and we also addressed the details of experimental design as follows.

Prior to P300 tasks, resting-state EEG data of four minutes was recorded. After 1-minute short break, the 3-times P300 tasks started. We measured 150 trials, including 120 standard stimuli and 30 target stimuli, in each time of P300 task. In each trial of P300, a bold cross appeared as a cue to remind subjects to focus their attention on the screen. 250 ms later, a thin cross lasting 500 ms appeared to inform subjects that a (target or standard) stimulus came out randomly. Then, a stimulus was presented to all subjects lasting 500 ms, and in the meantime, subjects were required to count the number of target stimuli. In our study, the target stimulus was the combination of the downward-oriented triangle with thin cross in the center, and the standard stimulus was the combination of the upward-oriented triangle with thin cross in the center. After 1-second break, a following P300 trial initiated. When one P300 loop completed, subjects were required to verbally speak out the total number of target stimuli which they had counted.

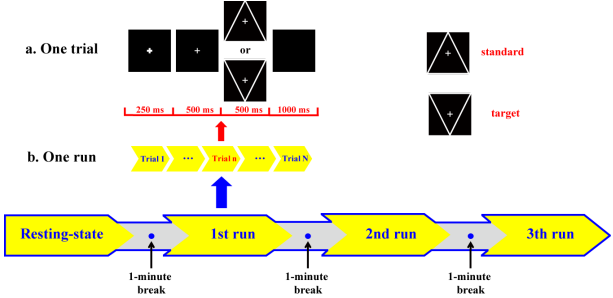

**FIGURE 1.** The P300 experimental protocol of our study. A 4 minutes resting-state and 3 times of P300 tasks were contained, and 1 minute break was designed between two conditions. The down-oriented triangle with a thin cross in the center indicated the target stimulus, and the upward-oriented triangle with a thin cross in the center represented the standard stimulus. In each run of P300 tasks, 150 trials were randomly presented. In each trial of P300, we included a 250-ms alert of attention, a 500-ms cue of preparation, a 500-ms target or standard stimulus, and a 1000-ms black screen.

### C. EEG DATA ACQUISITION

The EEG datasets were recorded with 64 Ag/AgCl electrodes that are positioned in compliance with the international 10/20 system and digitized with a sampling rate of 500 Hz (Brain Products GmbH) and an online band-pass filtering of 0.01-100 Hz. During EEG recordings, the electrodes FCz and AFz were regarded as the reference and ground, respectively. Meanwhile, the vertical and horizontal electrooculogram were separately recorded with 2 additional channels to monitor subjects' eye movements. Of note, we ensured that impedance of all electrodes was consistently maintained below $5\,k\Omega$.

### D. EEG DATA PROCESSES

To investigate the specified brain reconfiguration in different frequency bands from a resting-state condition to an oddball stimulus task, we first extracted the P300 amplitudes for all subjects, and then conducted functional network analysis to build the brain networks for both resting-state and stimulus task EEG signals. Finally, we investigated the underlying relationship between the brain reconfiguration in different frequency bands and P300 amplitudes.

### D.1. P300 AMPLITUDE

Before the calculations of P300 amplitudes, raw EEG signals were first preprocessed by using such procedures, including the reference electrode standardization technique (REST) re-referencing [46, 47], 1-13 Hz bandpass filtering, [-200 800] ms data segmentation (0 ms represents the onset of stimulus presentation), [-200 0] ms baseline correction, artifact-trial removing (±75 μv as the ocular threshold), and multiple P300 trial averaging. Subsequently, based on the averaged P300 ERP, the P300 amplitude was calculated within the time interval of [300 500] ms after the onset of stimulus. To access a reliable estimation of P300 amplitude [29], the mean of amplitudes was calculated across the five electrodes over posterior areas (i.e., CPz, CP1, CP2, Cz, and Pz) that have been demonstrated with significant P300 components. Of note, to reduce the side effect of noise, the P300 amplitude was finally defined as the averaged amplitude within the time window of ±10 ms with the largest positive peak at the center.

### D.2. TIME-FREQUENCY ANALYSIS

The time-frequency distributions (TFDs) using the wavelet analysis was conducted on Pz in the resting-state and P300 task, since the parietal area presented the greatest generator of P300 and highest P300 amplitude [48, 49]. Therefore, we first calculated the corresponding TFDs of each segment for each subject, and then acquired the mean values by averaging the TFDs across all segments. With the TFDs, we analyzed the fluctuations of the rhythmical activity between the resting-state condition and P300 task.

### D.3. BRAIN NETWORK

In this study, we used the canonical 21 electrodes of international 10-20 system [41]. The EEG procedures include the REST re-referencing, 1-30 band-pass filtering, and 2.25-seconds segmentation (an entire P300 trial). These resting-state (also P300 task) segments were then visually checked to remove the EEG segments contaminated with obvious ocular or head movement artifacts.

Coherence, a commonly used method in analyzing the cooperative and synchrony-defined neuronal assemblies, represents the linear relationship at a specific frequency between two signals. In our study, the coherence was used as a measure of the interactions between two electrodes in the EEG segments, $x(t)$ and $y(t)$, and coherence is expressed as follows:

$$C_{xy}(f) = \frac{|P_{xy}(f)|^2}{P_{xx}(f)P_{yy}(f)} \quad (1)$$

where $P_{xy}(f)$ denotes the cross-spectrum of $x(t)$ and $y(t)$ at frequency $f$, and $P_{xx}(f)$ and $P_{yy}(f)$ denote the related auto-spectrum of $x(t)$ and $y(t)$ at frequency $f$ estimated from the Welch-based spectrum, respectively.

The 21 electrodes were set as the network nodes, and the coherence value calculated between nodes was set as the network edge which was measured by averaging the $C_{xy}$ within the frequency band of interest. In our study, the 21×21 weighted adjacency matrix was thereby obtained for each EEG segment of each subject. Subsequently, the adjacency matrices were averaged across EEG segments to develop a final matrix for each subject under each brain condition.

To show the brain reconfiguration in different frequency bands, we first analyzed the topologies between resting and P300 task networks in marking the network edges which were significantly reconfigured (enhanced or suppressed), when the brain transferred from resting-state condition (i.e., baseline) to P300 task. In the meantime, to quantitatively measure the efficiency of reconfiguration from baseline to

P300 task, we then calculated the reconfigured coupling strengths by averaging the coupling difference of the significantly reconfigured network edges from baseline to P300 task.

### D.4. OUTLIERS REMOVING STRATEGY
During analyzing the underlying relationships between the P300 amplitudes and reconfigured coupling strengths, it exists some outlier subjects whose data obviously deviates from data center. In order to obtain a robust representative knowledge, the outlier subjects were discerned based on the Malahanobis distances [50]. Of them, subjects who had the largest 10% Malahanobis distances to the data center were excluded from the following analyses.

### D.5. CORRELATION ANALYSIS
In this study, Pearson's correlation was conducted to assess the two potential relationships: one index is the correlation between the two reconfigured coupling strengths in the different frequency bands, and the other index is the potential relationships between the P300 amplitudes and the brain reconfigurations of distinct rhythms.

### D.6. CLASSIFICATION OF HIGH- AND LOW-AMPLITUDE GROUPS
We finally explored the differences of the reconfigured coupling strengths between high- and low-amplitude groups. In specific, of all 21 subjects, nine subjects with highest and nine subjects with lowest amplitudes were included in the following analyses. Based on the mean of reconfigured strengths of the significantly increased/decreased edges, we first analyzed the statistical differences of the reconfigured coupling strengths between the two groups. And by taking the mean of the significantly reconfigured (increasing also decreasing) coupling strengths in the delta/theta and alpha bands as the discriminative features, we also evaluated the feasibility to differentiate the high- from low-amplitude group by using linear discriminant analysis and support vector machine with default parameters.

## III. RESULTS

### A. SPECIFIC RHYTHMICAL BRAIN RECONFIGURATION
For the specified brain activity during rest and task in the different rhythms, the TFDs in Figure 2 demonstrate that the alpha rhythm plays the dominated role in spontaneous brain activity of baseline condition, while brain activity in the alpha rhythm was then attenuated and suppressed when subjects started to participate in P300 tasks. In this progress, the task-related brain activity in frequency range of 1-8 Hz that covers the delta and theta rhythms were inversely enhanced to process the incoming information of P300.

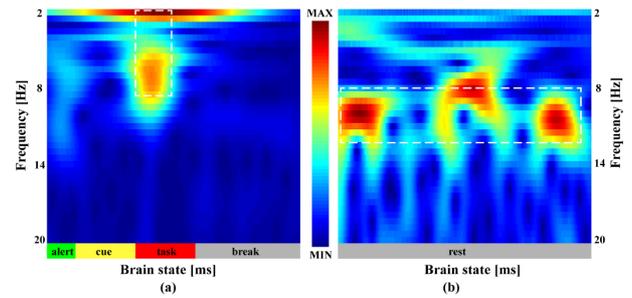

**FIGURE 2.** The dynamic TDFs of the rhythmical brain activity during rest and task. (a) Task state, (b) Rest state.

The differences of network topology that reflect the brain reconfiguration in frequency ranges of 1-8 and 8-13 Hz from baseline to P300 task are demonstrated in Figure 3. From which, we could see that within the frequency of 1-8 Hz, the coupling strengths between brain areas including frontal/prefrontal and parietal lobes are enhanced ($p < 0.05$, Bonferroni correction), when the brain switches from baseline to P300 task. In contrast, in the alpha rhythm, the decreased default mode network (DMN)-like topology appears within this progress ($p < 0.05$, Bonferroni correction).

In the brain, the increasing or decreasing of brain activity is balanced, since a marginally significantly ($p = 0.067$) negative relationship ($r = -0.407$) between the increased coupling strengths in frequency of 1-8 Hz and decreased coupling strengths in frequency of 8-13 Hz is demonstrated in our study.

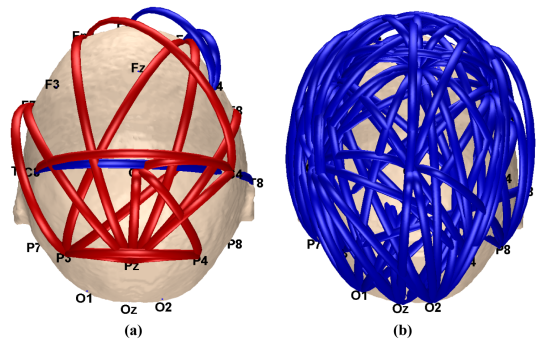

**FIGURE 3.** Brain network topology updates (increasing or decreasing) of the couplings from rest to task within the frequency bands of 1-8 Hz and 8-13 Hz. (a) 1-8 Hz, (b) 8-13 Hz, In (a) and (b), the red and blue solid lines indicate the increased and decreased network coupling strengths from rest to task, respectively.

### B. CORRELATIONS BETWEEN BRAIN RECONFIGURATION AND P300 AMPLITUDES
Since the reconfigured coupling strengths from baseline to P300 task were achieved for all subject, the relationships between the P300 amplitudes and reconfigured coupling strengths in frequency ranges of 1-8 and 8-13 Hz are shown separately in Figure 4. The results demonstrated that the strength mean of increased coupling in frequency of 1-8 Hz is positively ($p < 0.05$, Bonferroni correction) related to the P300 amplitudes ($r = 0.532$); in contrast, that of decreased linkages in the alpha band is negatively correlated with the P300 amplitudes ($r = -0.364$), although it could not pass the Bonferroni correction.

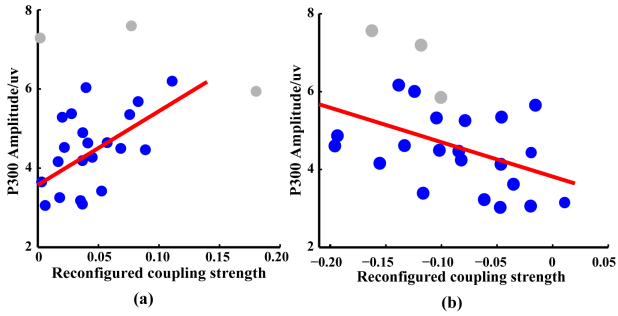

**FIGURE 4.** Correlations between the reconfigured coupling strengths and P300 amplitudes within the frequency ranges of 1-8 and 8-13 Hz. (a) 1-8 Hz, (b) 8-13 Hz. In each sub-figure, the blue filled circles denote the subjects' P300 amplitudes, the grey filled circles denote the excluded outlier subjects, and the red solid lines are the linear regression lines of the P300 amplitudes to the reconfigured coupling strengths.

### C. CLASSIFICATION BETWEEN LOW- AND HIGH-AMPLITUDE GROUPS

Figure 5 demonstrates the differences of reconfigured coupling strengths from baseline to P300 task conditions in frequency ranges of 1-8 and 8-13 Hz between the low- and high-amplitude groups. For subjects with high P300 amplitudes, the significantly ($p = 0.006$) larger increased coupling strengths in frequency range of 1-8 Hz and marginally significantly ($p = 0.051$) smaller decreased coupling strengths in the alpha band could be observed, when compared to that of low-amplitude subjects.

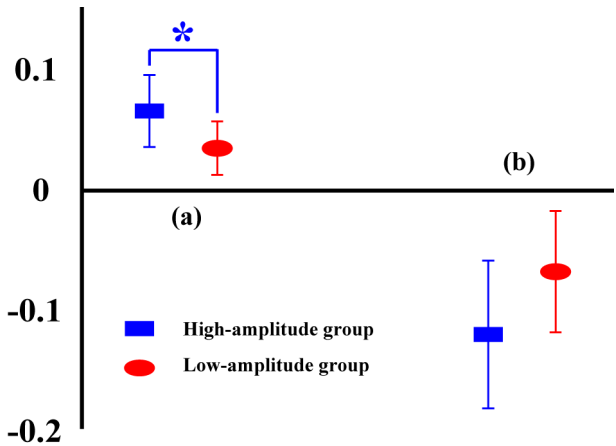

**FIGURE 5.** The reconfigured coupling strengths between high- and low-amplitude group within the frequency ranges of 1-8 Hz and 8-13 Hz. (a) 1-8 Hz, (b) 8-13 Hz. In each sub-figure, the blue and red bars indicate the high- and low-amplitude group, respectively, and the blue colored star indicate the significance with $p < 0.05$ between the two groups.

As demonstrated in Figure 6, the differentiated distributions of reconfigured coupling strengths between the high- and low-amplitude groups could be found in the progress of brain reconfiguration from baseline to P30 task conditions. Based on the different reconfiguration patterns, the classification analysis revealed that only four out of eighteen subjects were falsely classified into the opposite group, which achieved a classification accuracy of 77.78% to differentiate the two groups.

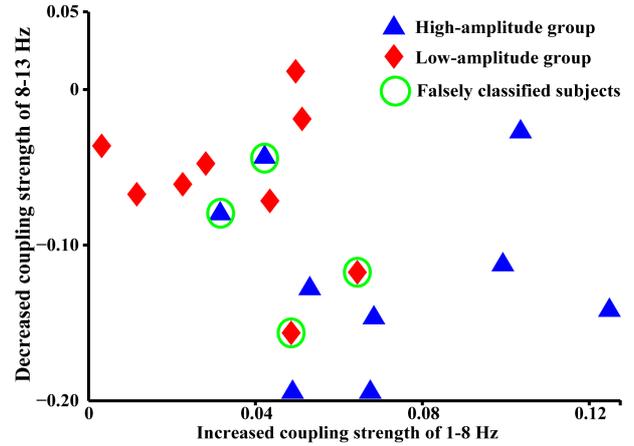

**FIGURE 6.** Scatter for the mean of reconfigured (decreased also increased) coupling strengths in frequency ranges of 8-13 and 1-8 Hz. The red diamonds denote the low-amplitude group, the blue triangles denote the high-amplitude group, and the green circles denote the falsely classified subjects.

### IV. Discussion

Previous studies have demonstrated that the specified brain activities in different rhythms (e.g., delta, theta, and alpha bands) significantly relate to the P300 components [18-21]. Since no target-related information is received by the brain, the alpha rhythm plays the dominant role in the spontaneous brain activity in resting state. Thus, the TFDs shown in Figure 2(b) demonstrate a larger magnitude in the alpha band spanned across the whole time duration. However, when subjects were required to perform the P300 tasks, much target-related information is needed to be processed in the brain. The brain thus involves such functions including attention, signal matching, and decision making to process the received information, which is accomplished by the specified brain activity in the delta and theta bands [18-20]. Herein, Figure 2(a) demonstrates the enhanced brain activity in the delta and theta bands during the P300 tasks. As a consequence, the brain activity in the alpha rhythm is attenuated and suppressed [19, 21] (i.e., alpha event-related desynchronization). We thus assumed that the specified reconfigured couplings from baseline to P300 task in different frequency bands are governed by the needs of information processing in the brain, and the differentiable network topology could thereby account for the underlying mechanism of the generations of P300.

In this study, the specified network topologies in different rhythms, which reflect the reallocation of limited brain resources, are given in Figure 3. The opposite reconfigured trends of specified brain activity that is related to P300 components are demonstrated in different frequency bands; in particular, the parameters quantitatively evaluating both the increased topologies in frequency of 1-8 Hz and decreased topologies in frequency of 8-13 Hz are further demonstrated to relate to the P300 components (Figure 4). The P300 is attributed to the involvements of such brain functions as attention and decision making [3, 51, 52] that are functioned on large-scale brain networks [30, 53-55]. As illustrated, the interactions between frontal (also prefrontal) and parietal lobes play the crucial roles in the generations of P300, and reductions of P300 amplitudes are observed in frontal-lobe lesioned patients [14]. In Figure 3, in frequency range of 1-8 Hz, the network edges between frontal and parietal lobes are enhanced (Figure 3(a)); while in the alpha band, such notion of alpha suppression as decreased DMN-like topology in this study is demonstrated, which has been also clarified by previous related studies [19, 21]. DMN is deactivated in tasks, and is thereby negatively correlated with experimental task [56, 57].

Our brain works daily in a balance way to effectively process information. A marginally significantly negative relationship between the increased coupling strengths in frequency of 1-8 Hz and decreased coupling strengths in the alpha band is demonstrated in this study, i.e., the stronger the negative network is configured in the alpha band, the stronger the positive network will be subsequently reconfigured in the delta/theta band. Which means, in task-related EEG rhythms, the interactions between frontal and parietal lobes, which are the main generators of P300 [12, 29, 49], are enhanced; whereas to compensate the increased task-related brain activity in frequency of 1-8 Hz, the brain activity in the alpha rhythm is inversely suppressed, which corresponds the decreased coupling strengths of DMN.

Besides the relationship between increased and decreased coupling strengths in different frequency bands, this study further demonstrated the close relationships between the reconfigured coupling strengths in different frequency bands and P300 amplitudes, which was depicted in Figure 4. Theoretically, the reconfigured coupling strengths represent quantitatively the adjustment degree of a brain network to fulfill the requirement for efficient information processing. The positive relationship (Figure 4a) of increased strengths versus P300 amplitudes and negative relationship (Figure 4b) of decreased strengths versus P300 amplitudes consistently demonstrate that a person who has the more efficient brain network reconfigured from baseline to P300 task conditions might have the greater potential to evoke a P300 component with the larger amplitude, when he (or she) participates in our P300 tasks.

In this study, this inference was finally investigated by implementing the comparison of reconfigured coupling strengths in frequency ranges of 1-8 and 8-13 Hz between high- and low-amplitude groups. Subjects with higher P300 amplitude are demonstrated to have the larger increased coupling strengths in frequency of 1-8 Hz and much smaller decreased coupling strengths in the alpha band (Figure 5); in the meantime, the classification between high- and low-amplitude groups shown in Figure 6 consistently clarified that the reconfigured coupling strengths in two concerned frequency bands indeed relate to the generations of P300.

## V. Conclusion

In our study, the dynamic network topologies were demonstrated to reconfigure from resting-state to stimulus task within different frequency bands (i.e., delta/theta and alpha bands) in an efficient and balanced manner. Furthermore, the reconfigured network topologies in different bands were clarified to closely relate to the P300 components evoked by the presentations of target stimuli in P300 tasks. Findings of our study indicated that the reconfigured network pattern could be potentially used for the selection of applicable users for P300-BCI, and also helpful to deepen our knowledge of the generations of P300.